\documentclass[fleqn,usenatbib]{mnras}
\usepackage{newtxtext,newtxmath}
\usepackage[T1]{fontenc}
\DeclareRobustCommand{\VAN}[3]{#2}
\let\VANthebibliography\thebibliography
\def\thebibliography{\DeclareRobustCommand{\VAN}[3]{##3}\VANthebibliography}
\usepackage{graphicx}	
\usepackage{amsmath}	
\usepackage{subcaption}

\title[Black Widow]{Polarization Studies of Black Widows PSRs B1957+20, J2055+3829 and J1544+4937}

\author[Wang et al.]{
S.Q. Wang$^{1,2,3}$, \thanks{E-mail: wangshuangqiang@xao.ac.cn}
N. Wang$^{1,3}$, S. Dai$^{2}$, G. Hobbs$^{2}$, R. Luo$^{4}$, J.B. Wang$^{5}$, and A. Zic$^{2}$
\\
$^{1}$Xinjiang Astronomical Observatory, Chinese Academy of Sciences, Urumqi, Xinjiang 830011, China \\
$^{2}$CSIRO Astronomy and Space Science, PO Box 76, Epping, NSW 1710, Australia\\
$^{3}$Key Laboratory of Radio Astronomy and Technology, Chinese Academy of Sciences, A20 Datun Road, Chaoyang District, Beijing, 100101, China\\
$^{4}$Department of Astronomy, School of Physics and Materials Science, Guangzhou University, Guangzhou 510006, China\\
$^{5}$Institute of Optoelectronic Technology, Lishui University, Lishui, Zhejiang, 323000, China\\
}

\date{Accepted XXX. Received YYY; in original form ZZZ}

\pubyear{\the\year{}}

\begin{document}
\label{firstpage}
\pagerange{\pageref{firstpage}--\pageref{lastpage}}
\maketitle

\begin{abstract} 
We present an analysis of the polarization  of three black widow pulsars, PSRs B1957+20, J2055+3829 and J1544+4937 at 1250\,MHz using the Five-hundred-meter Aperture Spherical radio Telescope (FAST). Radio eclipses for PSRs B1957+20 and J2055+3829 are detected, while the radio emission for J1544+4937 is detected throughout the eclipse. We study the polarization and dispersion measure properties of the pulsars near and during the eclipse. The position angle of the linear polarization is observed to shift at the eclipse boundary in all of these three pulsars implying a lower limit line-of-sight magnetic field strength of the eclipse medium of approximately a few mG to tens of mG. We also find evidence that the magnetic field in the eclipse medium of PSRs B1957+20 and J1544+4937 reverses. 
\end{abstract}




\section{Introduction}

\begin{figure*}
\centering
\includegraphics[width=55mm]{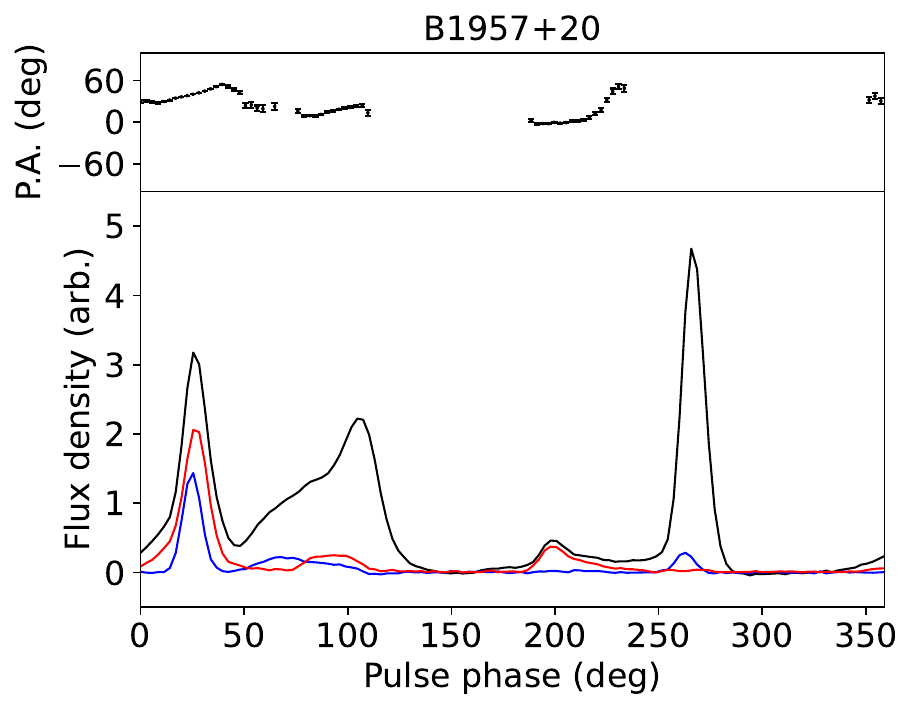}
\includegraphics[width=55mm]{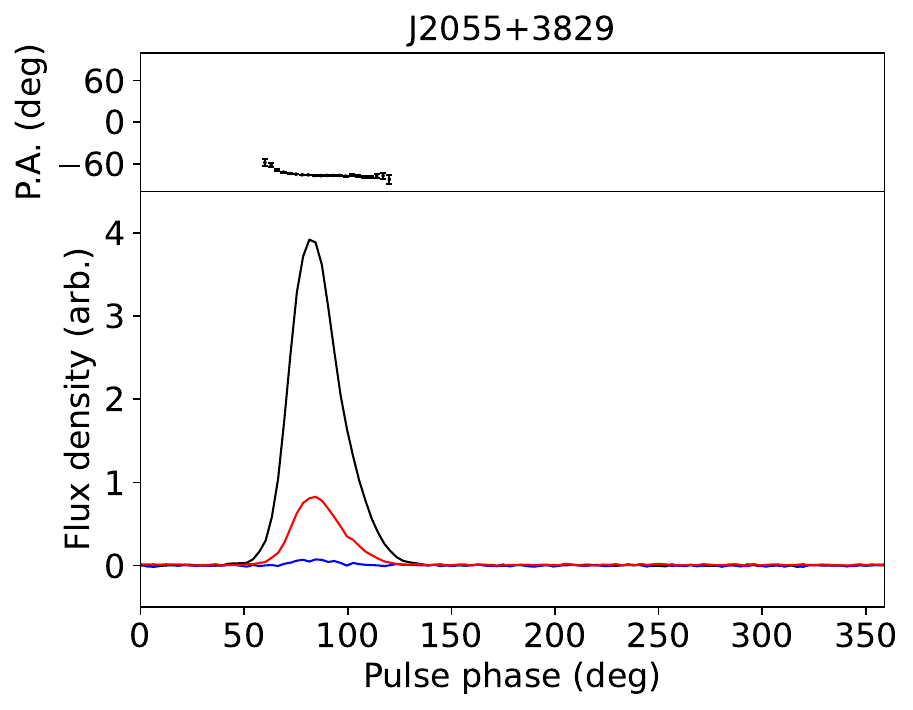}
\includegraphics[width=55mm]{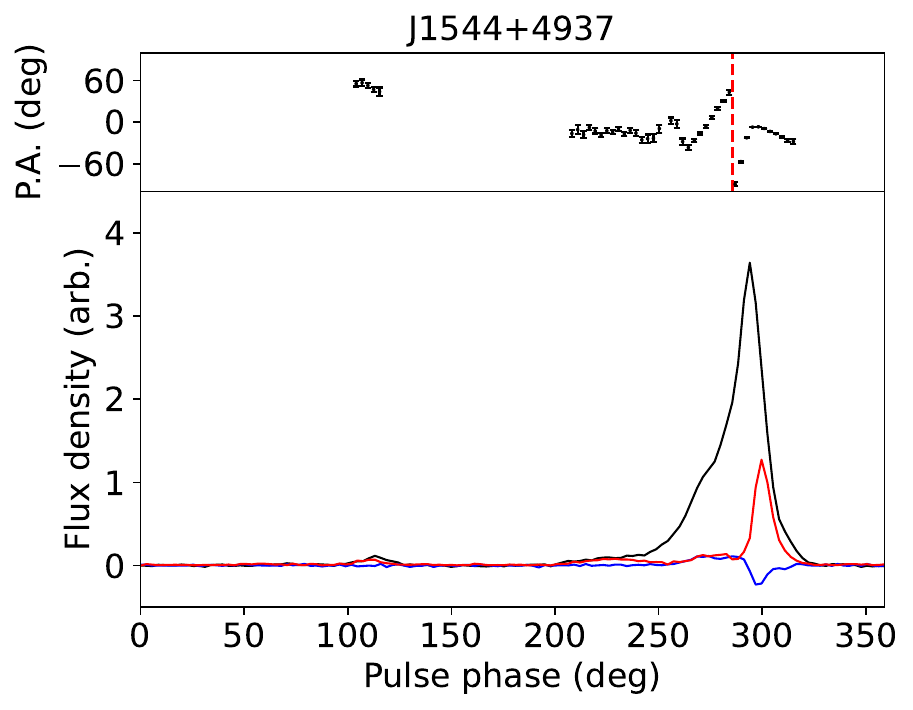}
\caption{Polarization profiles for PSRs B1957+20, J2055+3829 and J1544+4937. The black, red, and blue lines are for the total intensity, linear polarized intensity, and circular polarized intensity, respectively. The position angles (black dots) of the linear polarized emission are shown as black dots.}
\label{prof}
\end{figure*}

Spider pulsars are a special group of radio millisecond pulsars (MSPs) characterized by a low-mass companion and a short orbital period~\citep{Fruchter1988}. There are two types of spider pulsars: redbacks (RBs) with companion masses of approximately from 0.2 to 0.4 $M_{\odot}$ and black widows (BWs) with companion masses of approximately from 0.01 to 0.05 $M_{\odot}$~\citep{2013IAUS..291..127R}. In spider pulsars, the companion is evaporated by the pulsar winds and high-energy emissions, causing material to be blown from the companion, which can result in radio eclipses~\citep{Fruchter1988}. The study of radio eclipses is important for understanding the properties of the eclipse medium and the evolution of MSPs.

Observations over a wide frequency range reveal that the eclipse is frequency-dependent~\citep{Stappers2001a}. The eclipse generally exhibits a shorter duration or even disappears at higher observational frequencies. The eclipse mechanism for spider pulsars is unclear, and different observational frequencies may involve different eclipse mechanisms~\citep{Thompson1994}. Pulse scattering, pulse smearing at lower frequencies, or cyclotron damping may be the possible mechanisms causing the eclipse~\citep{Stappers2001a, Polzin2019, Bai2022}. Polarization analysis is critical for understanding the properties of the eclipse medium and the eclipse mechanism.

PSRs B1957+20, J2055+3829, and J1544+4937 are three BWs in the galactic disk. PSR B1957+20 is the first discovered BW with a 1.6 ms spin period and a 9.17 hr orbital period~\citep{Fruchter1988}. 
This system may harbor a massive pulsar with a mass of $2.40\pm0.12\,M_{\odot}$~\citep{Kerkwijk2011}. Eclipses have been observed over a wide frequency range, from approximately 100 MHz to 2 GHz~\citep{Fruchter1990, Ryba1991, Polzin2020}.
PSR J2055+3829 has a spin period of 2.1 ms and an orbital period of 3.1 hr, and it was discovered by the SPAN512 survey~\citep{Guillemot2019}. Using the Nançay Radio Telescope, \citet{Guillemot2019} measured the eclipse duration of the pulsar to be about 10\% of the orbit at 1.4 GHz. PSR J1544+4937 has a spin period of 2.2 ms and an orbital period of 2.9 hr, and it was discovered by the Giant Metrewave Radio Telescope (GMRT) by searching for radio pulsations in the directions of unidentified Fermi-Large Area Telescope $\gamma$-ray sources~\citep{Bhattacharyya2013}. This pulsar shows a clear eclipse at 322 MHz with an eclipse duration of 13\% of the orbital period, while it is detected throughout the eclipse at 607\,MHz~\citep{Bhattacharyya2013}.

The radio emissions of spider pulsars are generally weak, and studies of eclipses are limited by low signal-to-noise ratios (S/N). More sensitive observations of spider pulsars would be useful for characterizing the eclipse medium. In this paper, we present polarimetric observations of three BW pulsars PSRs B1957+20, J2055+3829, and J1544+4937, using the Five-hundred-meter Aperture Spherical radio Telescope (FAST). In Section 2, we describe our observations and data processing. In Section 3, we present the results. We discuss and summarize our results in Section 4.

\begin{figure*}
\begin{center}
\begin{tabular}{c}
\includegraphics[width=160mm]{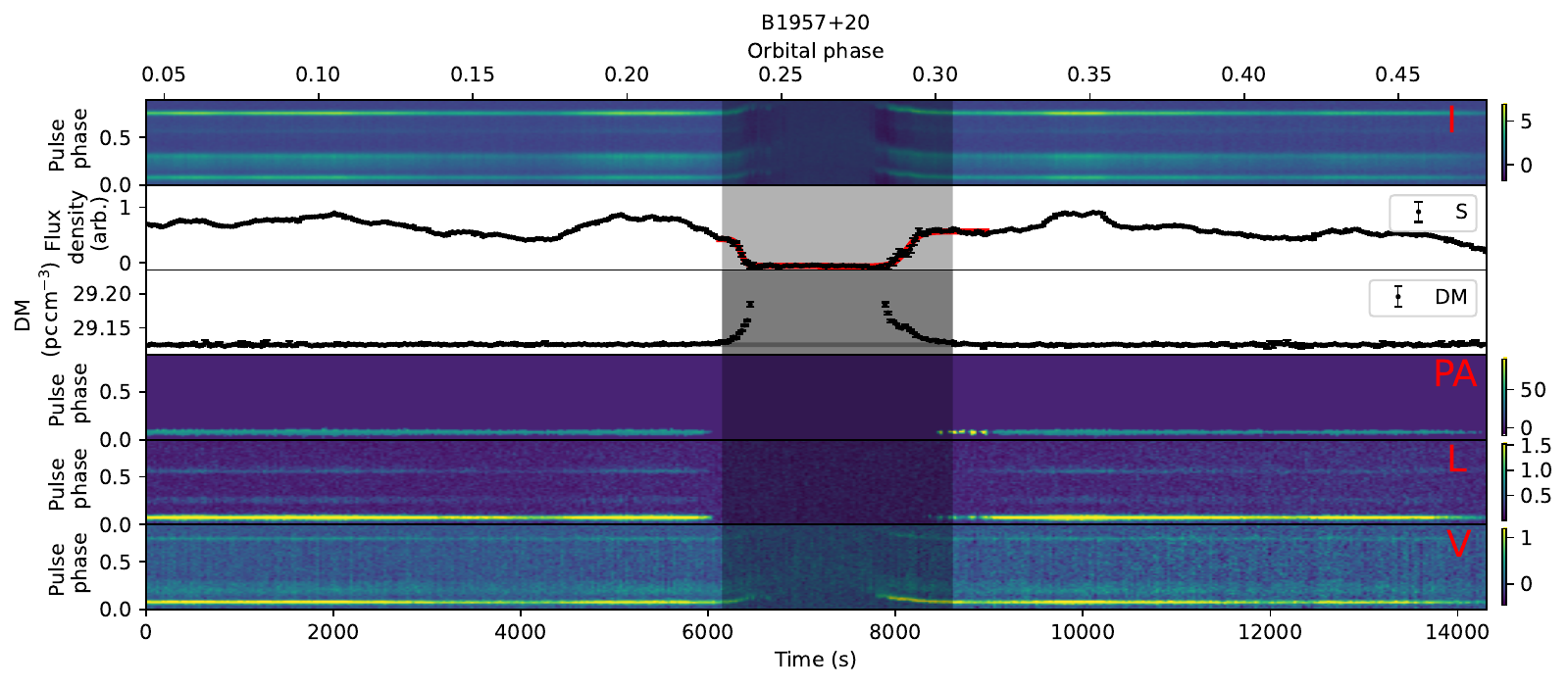}\\
\includegraphics[width=160mm]{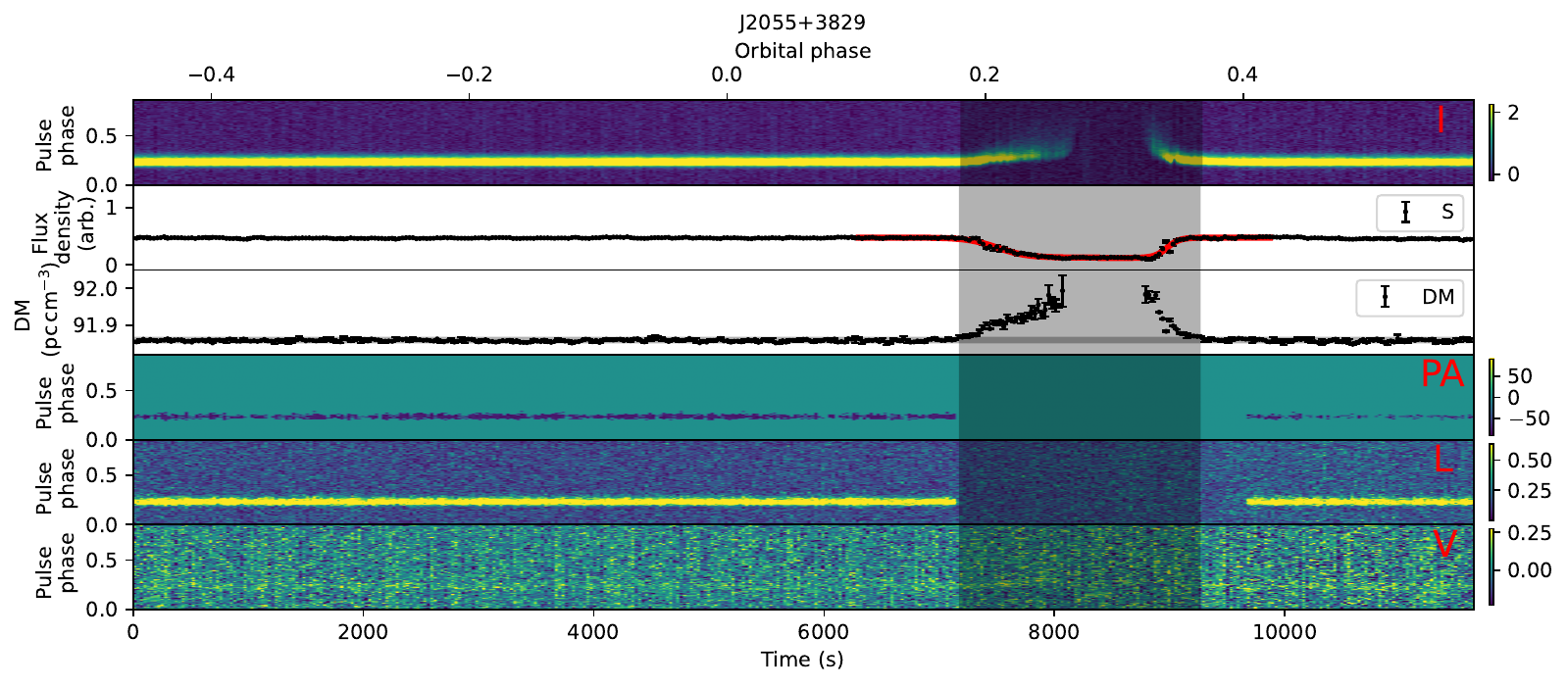}\\
\includegraphics[width=160mm]{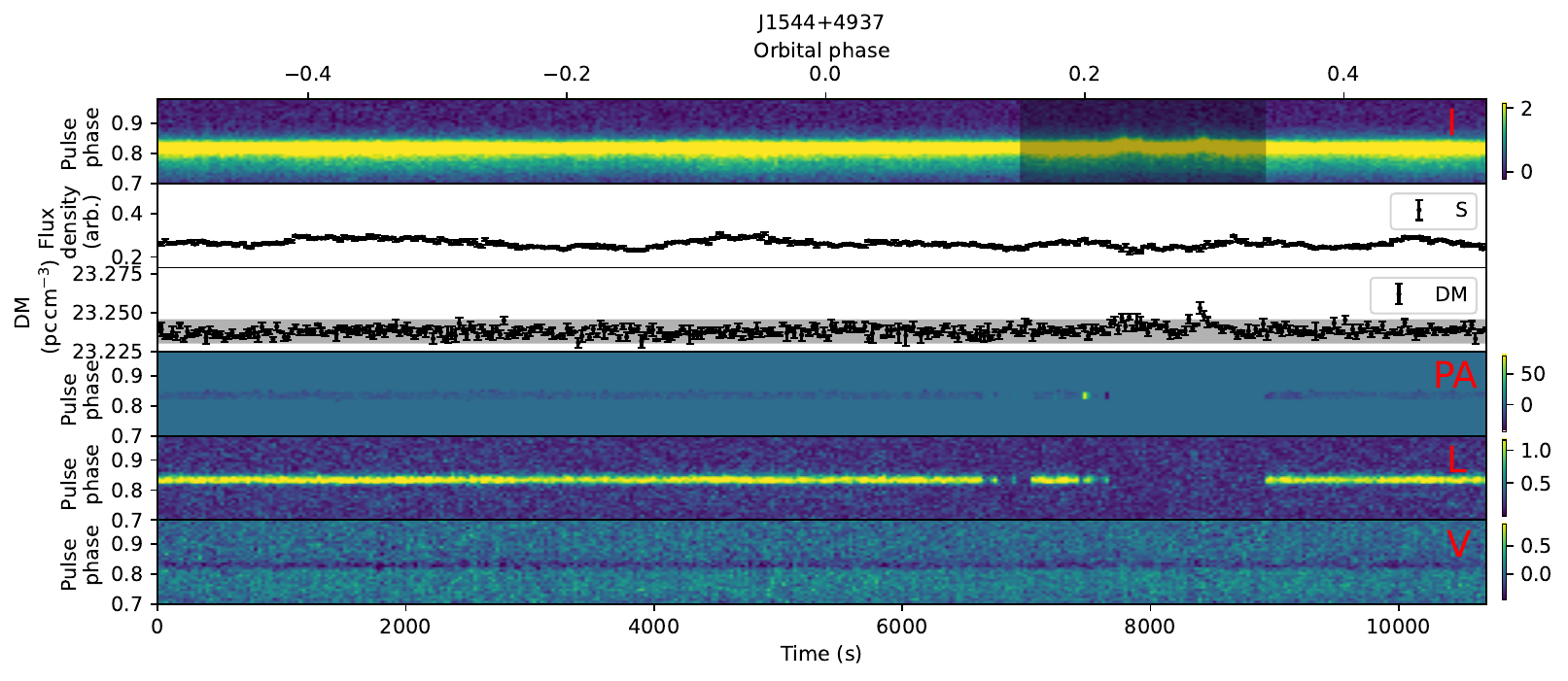}\\
\end{tabular}
\end{center}
\caption{The intensity, flux density, DM, RM, PA swings, linear and circular polarizations of PSRs B1957+20, J2055+3829, and J1544+4937 with a sub-integration of 30\,s versus orbital phase. The red solid lines in the second panel denotes the fitting of the Fermi-Dirac function. The horizontal filled grey area in the third panel identifies the region of 3$\sigma$ of DM variations. The vertical filled grey areas identify the region where DM shows significant variations.}
\label{pola}
\end{figure*}

\section{OBSERVATIONS AND DATA PROCESSING}

The observations of PSRs B1957+20, J2055+3829, and J1544+4937 were carried out on 2nd Sep. 2022, 30th Aug. 2021, and 18th Sep. 2022, respectively, using the central beam of the 19-beam receiver of FAST, with the frequency covering 1.05-1.45 GHz. The observational durations for PSRs B1957+20, J2055+3829, and J1544+4937 were 14340\,s, 11640\,s, and 10740\,s, respectively. The data were recorded in search mode PSRFITS format with four polarizations, 8-bit, 1024 frequency channels, and a 8.192\,$\mu$s sampling interval for PSRs B1957+20 and  J1544+4937, and a 16.384\,$\mu$s sampling interval for PSR J2055+3829.
For PSRs J2055+3829, and J1544+4937, our observations cover the entire binary orbital phases, but for PSR B1957+20, our observation only covers part of the binary orbital phase because of the limited tracking time of FAST. 
To analyze the dataset, we used {\sc dspsr}~\citep{Straten2011} to fold the data according to the timing ephemeris. Subsequently, {\sc paz} and {\sc pazi} in the {\sc psrchive} program~\citep{Hotan2004} were used to flag and remove narrow-band and impulsive radio-frequency interference (RFI). In our observations, a polarization calibration noise signal was recorded before or after the observation. We folded the calibration files, and then the pulsar observations were calibrated using {\sc pac} in the {\sc psrchive} program.

The rotation measure (RM) was measured using {\sc rmfit}, employing the brute-force search method to identify the peak in linear polarization. Our measurements of RM for PSRs B1957+20, J2055+3829, and J1544+4937 during out-of-eclipse phases are $-68.3\pm0.3$ ${\rm rad\,m^{-2}}$, $-63.7\pm0.5$ ${\rm rad\,m^{-2}}$, and $11.4\pm0.7$ ${\rm rad\,m^{-2}}$, respectively. The ionospheric contribution to RM for these pulsars was calculated to be $0.34\pm0.04$ ${\rm rad,m^{-2}}$, $0.12\pm0.04$ ${\rm rad,m^{-2}}$, and $1.51\pm0.05$ ${\rm rad,m^{-2}}$, respectively, using {\sc IonFR}~\citep{Sotomayor-Beltran2013}. Our RM measurements agree with previous results from \citet{Ng2020} and \citet{Spiewak2022}. 

The dispersion measure (DM) was measured using the {\sc tempo2} software package~\citep{Hobbs2006}. We divided the entire band into four equal sub-bands with a bandwidth of 100 MHz each. We formed a noise-free template by smoothing the integrated profile of the entire observation after removing the eclipse phase, and then times of arrival (ToAs) for each sub-integration were formed. The DM value of each sub-integration was obtained by fitting the ToAs at different sub-bands. The measured DM values for PSRs B1957+20, J2055+3829, and J1544+4937 in the out-of-eclipse region are 29.1247$\pm$0.0004 ${\rm cm^{-3}\,pc}$, 91.880$\pm$0.002 ${\rm cm^{-3}\,pc}$, and 23.239$\pm$0.002 ${\rm cm^{-3}\,pc}$, respectively.

To measure the eclipse duration, we used the Fermi-Dirac function to fit the flux densities during the ingress and egress of the eclipse of each pulsar according to the method of \citet{Polzin2019}. The Fermi-Dirac function is $f=a/(e^{{(t +p_1)}/{p_2}}+1)$, where $a$ is the amplitude, $t$ is the time, and $p_1$ and $p_2$ are fitted free parameters. The duration of the eclipse is taken to be the full width at half-maximum of the flux density. The duration of the ingress/egress of the eclipse is determined as the deviation between the orbital phases at 90\% and 10\% of the out-of-eclipse flux density.

\section{RESULTS}

\subsection{Polarization profile}

The polarization profiles of PSRs B1957+20, J2055+3829, and J1544+4937 are shown in Figure~\ref{prof}. PSR B1957+20 exhibits interpulse emission, and both the main pulse and interpulse show two components. The position angle (PA) swings exhibit smooth variations. PSR J2055+3829 has a simple single peak profile with smoothed PA swings. For PSR J1544+4937, we detected a weak interpulse with the peak intensity about 3\% of the main pulse. There is a jump of approximately 130 degree in the PA swings at the pulse phase of 285 degrees (indicated by the red vertical dashed line).

\subsection{Eclipse}

\subsubsection{PSR B1957+20}

\begin{figure*}
\centering
\includegraphics[width=150mm]{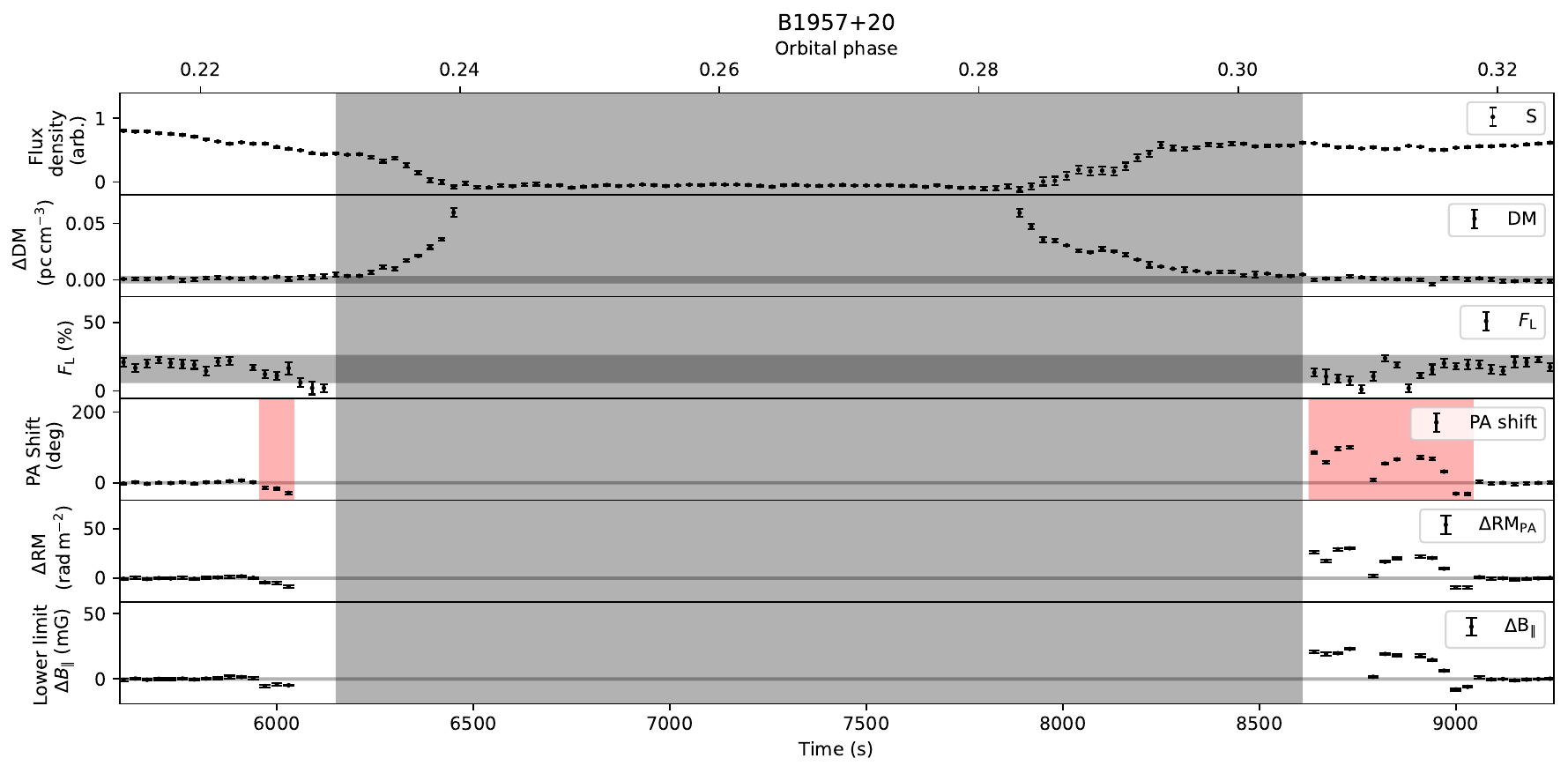}
\includegraphics[width=150mm]{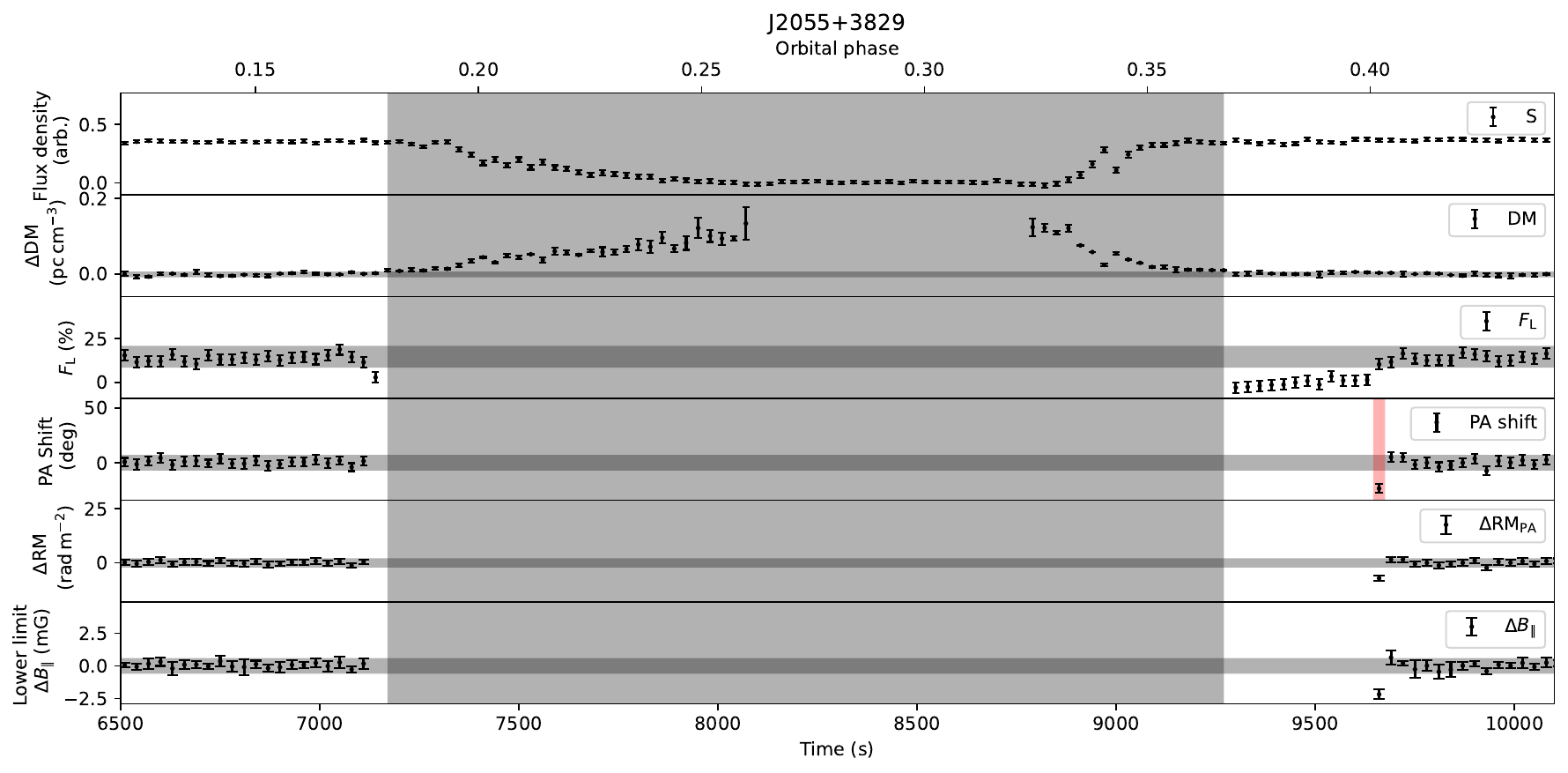}
\includegraphics[width=150mm]{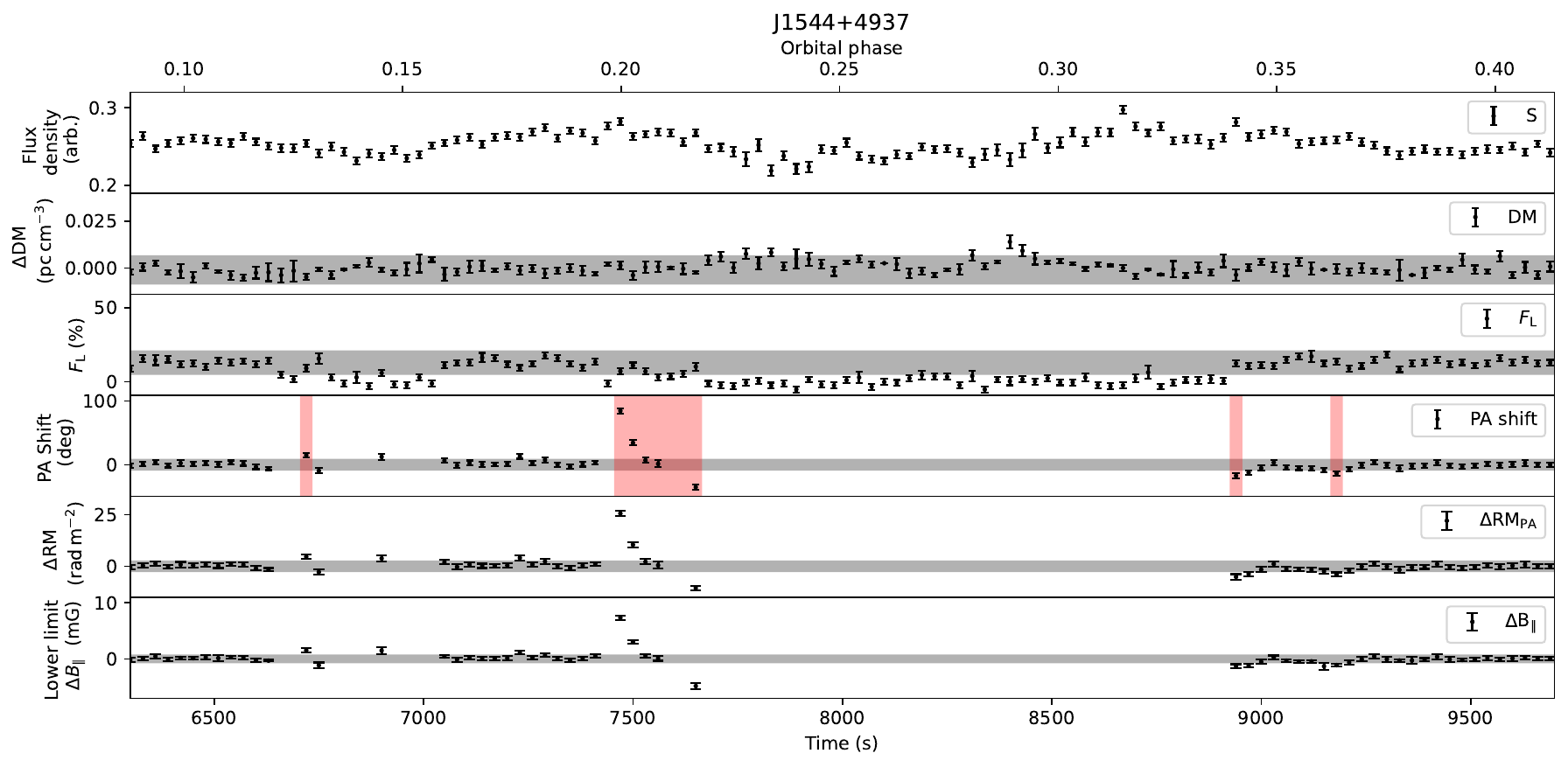}
\caption{The flux density, $\Delta$DM, linear polarization fraction ($F_{\rm L}$), degrees of PA shift ($\Delta$PA), $\Delta$RM, and the estimated lower limit $\Delta B_{\|}$ of PSRs B1957+20, J2055+3829, and J1544+4937 versus orbital phase. The horizontal filled grey area in each panel identifies the region of 3$\sigma$, while the vertical filled grey areas identify the orbital phases where the DM shows significant variations. The vertical filled red areas in the fourth panel identify the orbital phases where significant PA shifts are detected. Note that we used the uncertainty in $\Delta\mathrm{DM}$ to estimate $\Delta B_{\|}$, therefore, the actual shape of $\Delta B_{\|}$ remains uncertain.}
\label{pa}
\end{figure*}

The intensity, flux density, DM, RM, and polarization properties of PSR B1957+20 with a sub-integration of 30\,s versus orbital phase are shown in Figure \ref{pola}. A clear eclipse is detected for PSR B1957+20, during which the flux density decreases and subsequently increases. The flux density outside the eclipse is highly unstable due to scintillation. The eclipse duration is 1784$\pm$12\,s, approximately 5\% of the orbital period, with ingress and egress durations of 137$\pm$12\,s and 335$\pm$12\,s, respectively.  
Our results generally agree with the previous measurement of 1980$\pm$480\,s at 1400\,MHz reported by \citet{Ryba1991}.
The longer egress duration may be attributed to a cometary-like tail of outflowing material from the companion, which is shocked by the pulsar wind.

The DM shows significant increases during the eclipse, with the orbital phase ranging from 0.231 to 0.305 (the vertical filled grey area in Figure~\ref{pola}). During the out-of-eclipse phase, the DM remains unchanged with variations less than 3$\sigma$ of 0.003\,$\rm cm^{-3}\,pc$ (the horizontal filled grey area in Figure~\ref{pola}), where $\sigma$ is calculated by fitting the distribution of DM values of sub-integrations in the out-of-eclipse phase using the Gaussian function. The maximum DM increase during ingress and egress is about 0.06\,$\rm cm^{-3}\,pc$ compared to the out-of-eclipse DM of 29.1247$\pm$0.0004\,$\rm cm^{-3}\,pc$. There are structures in the DM variations during ingress and egress, as well as in the flux density, suggesting that the eclipse medium is highly clumpy~\citep{Tavani1991}. The DM increase rates during ingress and egress are 0.0002\,$\rm cm^{-3}\,pc\,s^{-1}$ and 0.00008\,$\rm cm^{-3}\,pc\,s^{-1}$, respectively, indicating a smaller particle density gradient in the cometary-like tail than in the head.

To study the polarization properties of PSR B1957+20, an RM value of $-68.3\pm0.3$ ${\rm rad,m^{-2}}$, derived from the average profile of the out-of-eclipse phase, was applied to correct the profiles of all sub-integrations. The PA swings of each sub-integration as a function of orbital phase are presented in the fourth panel of Figure~\ref{pola}.
If no magnetic field were present in the eclipse medium, the PA swings would remain consistent across all sub-integrations. However, we observe significant variations in the PA swings at the eclipse boundary, indicating the presence of a substantial line-of-sight magnetic field in the eclipse medium. 
Following the method of~\citet{Crowter2020}, we use the PA swings of the average profile from the out-of-eclipse phase as the reference PA swings.  
The degree of PA shift ($\Delta \rm PA$) for each sub-integration is then obtained by subtracting the reference PA swings (the first panel of Figure~\ref{pa}).  
Significant PA shifts are detected during the orbital phases $0.227-0.231$ and $0.305-0.318$ (marked by the filled red area in the first panel of Figure~\ref{pa}).  
The orbital phases between 0.231 and 0.305 are excluded from our $\Delta \rm PA$ analysis because significant DM variations and pulse scattering are observed in this range (the horizontal filled grey area in Figure~\ref{pola}), which could affect PA shift measurements.  
During the orbital phase $0.227-0.231$, $\Delta \rm PA$ exhibits a regular decrease, whereas during the orbital phase $0.305-0.318$, it shows more complex variations with multiple decreases and increases.  
The relationship between PA and RM is given by:  
\begin{equation}
\label{ep}
\Delta \rm PA = \Delta \rm RM \, \lambda ^2.  
\end{equation}
Using Equation~\ref{ep}, we calculate the $\Delta \rm RM$ for each sub-integration at different orbital phases (second panel of Figure~\ref{pa}).  
During the orbital phases $0.227-0.231$ and $0.305-0.318$, the $\Delta \rm PA$ values range from $-31\pm3$\,deg to $99\pm3$\,deg, corresponding to $\Delta \rm RM$ values of $-9\pm 1\,\rm rad\,m^{-2}$ to $30\pm1\,\rm rad\,m^{-2}$.  

We then use $\Delta \rm RM$ to estimate the line-of-sight magnetic field strength of the eclipse medium, given by:  
\begin{equation}
\label{e1}
\Delta B_{\parallel} = 1.23\mu G \frac{\Delta RM}{\Delta DM}.
\end{equation}
Since the DM does not exhibit significant variations during the orbital phases $0.227-0.231$ and $0.305-0.318$, we take the DM uncertainty for each sub-integration as $\Delta \rm DM$.  
The estimated lower limit $\Delta B_{\parallel}$ values, shown in the last panel of Figure~\ref{pa}, range from $-8.0\pm0.9$\,mG to $22.9\pm0.8$\,mG.  
Our results provide the first direct measurement of the line-of-sight magnetic field strength in the eclipse medium of PSR B1957+20.  
The opposite signs of the measured $ \Delta \rm RM$ offer evidence for magnetic field reversal in the eclipse medium.  

We then recorrect the profile of each sub-integration using the newly measured RM values.
For each sub-integration, if a PA shift is detected, the profile is corrected using the corresponding RM value obtained from the PA shift. Otherwise, the RM value from the out-of-eclipse phase is applied.
The linear and circular polarizations as functions of orbital phase are shown in the fifth and sixth panels of Figure~\ref{pola}, respectively.  
Since pulsar scattering and DM variations during the eclipse can alter polarization properties, we also exclude orbital phases with significant DM variations from our polarization analysis (the horizontal filled grey area in Figure~\ref{pola}).  
We find that linear polarization disappears during ingress of eclipse, while show some decreases during egress of eclipse, whereas circular polarization remains unchanged at these orbital phases. 
Given the complex magnetoionic environment of the eclipse medium, \citet{You2018} suggested that the depolarization could result from rapid RM variations within a sub-integration at the eclipse boundary. These variations arise due to small-scale fluctuations in both the density and the parallel component of the magnetic field within the eclipse medium.  
Assuming that RM variations within a sub-integration follow a normal distribution with a standard deviation $\sigma_{\rm RM}$, the depolarization due to RM variations can be described by~\citep{You2018}:  
\begin{equation}
\label{e2}
\frac{L}{ L_0} = \exp(-2\lambda^4\sigma_{\rm RM}^2),
\end{equation}
where $L_0$ and $L$ represent the linear polarization magnitude before and after integration over the fluctuations, respectively, $\sigma_{\rm RM}$ is the standard deviation of RM, and $\lambda$ is the wavelength.  
For PSR B1957+20, the fractional linear polarization is 21.5\% outside the eclipse.  
At the eclipse boundary, the linear polarization shows some decreases, even disappears, which may be corresponding to the different values of $\sigma_{\rm RM}$. {We arbitrarily assume that the fractional linear polarization decreases to less than 5\%.}  
Setting $L_0=21.5\%$, $L={5\%}$, and $\lambda=24\,{\rm cm}$, we estimate $\sigma_{\rm RM}\sim {15}\,\rm rad\,m^{-2}$.  
Thus, if $\sigma_{\rm RM}$ within a sub-integration (30\,s in our analysis) at the eclipse boundary exceeds ${15}\,\rm rad\,m^{-2}$, the fractional linear polarization can decrease to less than {5}\%.

\subsubsection{PSR J2055+3829}

The intensity, flux density, DM, RM, and polarization properties of PSR J2055+3829 with a sub-integration of 30\,s versus orbital phase are shown in Figure \ref{pola}, where a clear eclipse is detected. The eclipse duration is 1480$\pm$19\,s, about 13\% of the orbital period at 1250\,MHz, which is larger than the $\sim$10\% duration at 1.4 GHz reported by~\citet{Guillemot2019}. The durations of eclipse ingress and egress are approximately 568$\pm$10\,s and 237$\pm$9\,s, respectively. Unlike PSR B1957+20, the ingress lasts longer than the egress, suggesting that the companion's outflowing material maybe swept back.

The DM exhibits regular increases during the eclipse, with the orbital phase ranging from 0.180 to 0.367 (the filled grey area in the third panel of Figure~\ref{pola}). During the out-of-eclipse phase, the DM does not show significant variations within $3\sigma$ of 0.01\,$\rm cm^{-3}\,pc$ (the horizontal filled grey area in the third panel of Figure~\ref{pola}). Using the average profile of the out-of-eclipse phase, we measured the DM of the out-of-eclipse phase to be 91.860$\pm$0.002\,$\rm cm^{-3}\,pc$. The maximum DM during ingress and egress are 91.945$\pm$0.005\,$\rm cm^{-3}\,pc$ and 91.918$\pm$0.002\,$\rm cm^{-3}\,pc$, respectively, with corresponding DM variations of 0.085\,$\rm cm^{-3}\,pc$ and 0.058\,$\rm cm^{-3}\,pc$, respectively, compared to the out-of-eclipse phase. Unlike PSR B1957+20, the DM change rate during ingress is about $0.0002\,\rm cm^{-3}\,pc\,s^{-1}$, which is smaller than that during egress at $0.0003\,\rm cm^{-3}\,pc\,s^{-1}$.

Using the same method as for PSR B1957+20, the $\Delta \rm PA$ for each sub-integration is shown in the first panel of Figure~\ref{pa}, respectively.  
We detect a significant PA shift at the orbital phase of 0.402 (marked by the filled red area in the first panel of Figure~\ref{pa}), with a value of $-23\pm4$\,deg.  
Applying Equation~\ref{ep} and Equation~\ref{e1}, we obtain the corresponding values of $\Delta \rm RM= -7\pm1 \,\rm rad\,m^{-2}$ and {lower limit} $\Delta B_{\parallel} = -2.2\pm 0.3$\,mG.  

We then correct the profiles of sub-integrations using the newly measured RM values, following the method applied to PSR B1957+20.
Similar to the case of PSR B1957+20, the linear polarization of PSR J2055+3829 depolarizes before any noticeable changes in both DM and pulse profile, whereas the circular polarization seems remain unchanged at these orbital phases (the fifth and sixth panels of Figure~\ref{pola}).  
We assume that the depolarization of linear polarization can be modeled using Equation~\ref{e2}.  
{The fractional linear polarization of PSR J2055+3829 is 6\% outside the eclipse.}
By setting $L_0=6\%$, ${L=5\%}$, and $\lambda=24\,{\rm cm}$ in Equation~\ref{e2}, we obtain $\sigma_{\rm RM} \sim {5}\,\rm rad\,m^{-2}$.  
{Note that the  linear polarization fraction of PSR J2055+3829 is low and the significant PA shift is detected in only one sub-integration. Further observations are necessary to confirm our results. }

\subsubsection{PSR J1544+4937}

For PSR J1544+4937, an eclipse is detected at 322 MHz with the orbital phase ranging from $\sim$0.15 to 0.34, while emission is detected throughout the eclipse at 607 MHz~\citep{Bhattacharyya2013}. The intensity, flux density, DM, RM, and polarization properties of PSR J1544+4937 in our observation at 1250 MHz with a sub-integration of 30\,s versus orbital phase are shown in Figure~\ref{pola}. As expected, radio emission is detected throughout the entire orbital phase. The eclipse phase at 322 MHz of 0.15$-$0.34 is shown as the filled grey area in the first panel of Figure~\ref{pola}.

The flux density does not exhibit significant variations during the expected eclipse (the second panel of Figure~\ref{pola}). The DM during eclipse  exhibits significant increases only at the orbital phase of 0.289, while the DM variations remain within 3$\sigma$ of 0.008\,$\rm cm^{-3}\,pc$ (the filled grey area in the third panel of Figure~\ref{pola}) during the out-of-eclipse phase. The measured DM at the out-of-eclipse phase is $23.239\pm0.002\,\rm cm^{-3}\,pc$.

PSR J1544+4937 provides a unique opportunity to study the magnetic field within the eclipse medium, as its emission is detected throughout the eclipse.  
Using the same method as for PSR B1957+20, we present the $\Delta \rm PA$ for each sub-integration in the first panel of Figure~\ref{pa}.  
Significant $\Delta \rm PA$ variations are detected at certain orbital phases, such as 0.128, 0.341, and 0.364 (highlighted by the filled red areas in the first panel of Figure~\ref{pa}).  
In particular, during the orbital phases between 0.200 and 0.217, we observe a systematic decrease in $\Delta \rm PA$, from $84\pm4$\,deg to $-34\pm4$\,deg.  
Using Equation~\ref{ep}, this corresponds to a decrease in $\Delta \rm RM$ from $25\pm1\,\rm rad\,m^{-2}$ to $-10\pm1\,\rm rad\,m^{-2}$.  
Applying Equation~\ref{e1}, we determine that {lower limit} $\Delta B_{\parallel}$ decreases from $7.3\pm0.3$\,mG to $-4.8\pm0.5$\,mG.  
Similar to PSR B1957+20, the opposite signs of $\Delta \rm RM$ provide evidence for a magnetic field reversal within the eclipse medium.  

We then correct the profiles of sub-integrations using the newly measured RM values, following the method applied to PSR B1957+20.
During the expected eclipse phase of 0.15$-$0.34, the linear polarization is not consistently depolarized, suggesting a complex magnetoionic environment within the eclipse medium.  
For instance, the linear polarization depolarizes at the orbital phases of 0.120$-$0.158 and 0.195$-$0.348, while it remains unchanged at the orbital phase of 0.158$-$0.220.  
The circular polarization, however, remains unaffected throughout the eclipse.  
Similar to the case of PSR B1957+20, we assume that the decrease in linear polarization can be modeled using Equation~\ref{e2}.  
The fractional linear polarization of PSR J1544+4937 is 16\% outside the eclipse. 
By setting $L_0=16\%$, $L={5}\%$, and $\lambda=24\,{\rm cm}$ in Equation~\ref{e2}, we obtain $\sigma_{\rm RM}\sim {13}\,\rm rad\,m^{-2}$.

\section{ DISCUSSION AND CONCLUSIONS}

\subsection{Magnetic field in the eclipse medium and eclipse mechanism}

We present polarimetric observations of PSRs B1957+20, J2055+3829, J1544+4937 to constrain the magnetic field strength of the eclipse medium. For all three pulsars, we detect PA shifts at the eclipse boundary, with the estimated {lower limit} line-of-sight magnetic field strengths ranging from a few mG to several tens of mG. For both PSRs B1957+20 and J1544+4937, the sign of the $\Delta$RM changes, indicating the presence of magnetic field reversals in the eclipse medium. These results suggest that the magnetic field structure in the eclipse medium of spider pulsars is complex.

The strength of the perpendicular magnetic field in the eclipse medium can be constrained by analyzing variations in circular polarization. In the presence of a strong perpendicular magnetic field, the phase difference between X and O modes varies across the observing frequency, which can lead to a reduction in circular polarization~\citep{Thompson1994, Li2019}. However, for all these three pulsars, the circular polarization remains unchanged at the eclipse boundary where the linear polarization decreases. This suggests that the phase difference variation across the observed band is less than one full rotation. The upper limit of the perpendicular magnetic field strength can be estimated using the relation~\citep{Polzin2019}:
\begin{equation} \frac{B_{\perp}}{1{\rm G}} \leq \frac{1}{2.8} \left[ k_{\rm DM} \Delta {\rm DM} \left(\frac{1}{f^3_{\rm low}}-\frac{1}{f^3_{\rm high}}\right) \right]^{-1/2}, \end{equation}
where $k_{\rm DM}=e^2/2\pi m_e c \approx 4149\,{\rm s\, MHz^2 \,cm^3\,pc^{-1}}$, and $f$ is the observing frequency.
For our observations, $f_{\rm low}=1050$\,MHz and $f_{\rm high}=1450$\,MHz.
By taking the DM uncertainty for each pulsar as $\Delta \rm DM$, we estimate upper limits of $B_{\perp}$ as $\sim12$\,G for PSR B1957+20, $\sim5$\,G for PSR J2055+3829, and $\sim5$\,G for PSR J1544+4937.

Although the eclipse mechanism of spider pulsars remains unclear, previous studies suggest that pulse scattering or cyclotron damping could be responsible for the eclipses at L-band, e.g., \citet{Stappers2001a}.
Here, we primarily focus on the possibility of cyclotron damping, which requires a strong magnetic field in the eclipse medium. The relationship between the frequency of damped radiation and the magnetic field strength in the eclipse medium is given by~\citep{Polzin2019}:
\begin{equation} \frac{\nu_{\rm d}}{1\,{\rm GHz}} \approx 2.8\times10^{-3} \frac{B}{\gamma_{\rm p}(1-\cos \theta)}, \end{equation}
where $\nu_{\rm d}$ is the frequency of damped radiation, $\gamma_{\rm p}$ is the Lorentz factor of particles in the eclipse medium, and $\theta$ is the angle between the wave propagation direction and the magnetic field.
{Assuming $\gamma_{\rm p}(1-\cos \theta) \sim 1$, we find that $B \sim 450\,$G for $\nu_{\rm d}=$ 1.25\,GHz.

The magnetic field strength of the eclipse medium has been estimated for several other spider pulsars. For example, PSR J2256$-$1024 has a line-of-sight magnetic field strength of 1.11mG at an orbital phase of 0.32, with a binary separation of 0.08 light-seconds~\citep{Crowter2020}. PSR B1744$-$24A exhibits a lower limit magnetic field strength of 10 G at orbital phase 0.25 and a binary separation of 0.12 light-seconds~\citep{Li2023}. PSR J2051$-$0827 shows a lower limit line-of-sight magnetic field strength of 0.1 G at orbital phase 0.28 with a binary separation of 0.04 light-seconds~\citep{Wang2023}. Given that our magnetic field strength estimates represent lower limits, cyclotron damping cannot be ruled out. However, \citet{Li2019} constrained the magnetic field strength of the eclipse medium in PSR B1957+20 through plasma lensing and found that the cyclotron frequency is significantly lower than the observing frequency. This suggests that cyclotron damping is unlikely to be the dominant eclipse mechanism at L-band.
}

\subsection{The Depolarization Phenomenon}

We found that the depolarization phenomenon of linear polarization at the eclipse boundary in PSRs B1957+20, J2055+3829, and J1544+4937. This depolarization phenomenon has been observed in many other spider pulsars, such as PSRs J1748$-$2446A~\citep{You2018}, J1720$-$0533~\citep{Wang2021}, and J2051$-$0827~\citep{Wang2023}. The decrease in linear polarization may result from rapid RM variations within a sub-integration, driven by fluctuations in both the plasma density and the parallel component of the magnetic field in the eclipse medium~\citep{You2018}. Our observations show that linear polarization decreases before any significant variation in DM, suggesting that fluctuations in the line-of-sight magnetic field are the primary cause of depolarization.

Depolarization has also been observed in fast radio bursts (FRBs) (e.g., \citealt{2022Sci...375.1266F}), but in contrast to BWs, it occurs within individual bursts.
{The FRB depolarization is most likely attributed to multi-path propagation effects, while inter-pulse RM variations can be substantial in the case of pulsars.}
\citet{2025ApJ...982..154N} analyzed the polarization properties of 28 repeating FRBs and proposed that they can be classified into two categories: those originating from a stable magnetoionic environment and those associated with an evolving RM environment. By comparing RM variations between repeaters and pulsars, they found that FRB repeaters in evolving RM environments exhibit characteristics similar to BWs.
\citet{2022Sci...375.1266F} investigated depolarization in seven FRB repeaters and attributed it to the complex environments surrounding the bursts. They measured $\sigma_{\rm RM}$ values ranging from $0.12$ to $218.9\,\rm rad\,m^{-2}$. In comparison, we measured $\sigma_{\rm RM}$ values of $13$, $16$, and $20\,\rm rad\,m^{-2}$ for PSRs B1957+20, J2055+3829, and J1544+4937, respectively, comparable to those observed in FRB repeaters.
Given the intricate magnetic field structures in the eclipse medium of BWs, depolarization should theoretically be detectable in individual bursts. However, individual BW pulses are typically too weak to be observed. Notably, bright individual pulses, likely caused by strong lensing from intrabinary material, have been detected at the eclipse boundary of the BW PSR B1744$-$24A~\citep{Bilous2019}. Further polarization observations of these bright individual pulses at a wide range of frequencies (e.g., the Parkes radio telescope~\citep{Hobbs2020} and the upcoming Qitai radio telescope~\citep{2023SCPMA..6689512W}) will provide valuable insights for understanding the complex environments associated with both BWs and FRBs.

\subsection{Eclipse Duration}

In spider pulsars, the outflowing material from the companion is shocked by the pulsar wind, forming a cometary-like tail due to the binary's orbital motion~\citep{Fruchter1990}. As a result, the ingress duration is typically shorter than the egress duration.
In our observations, PSR B1957+20 follows this expected trend, with a shorter ingress duration compared to the egress duration, as well as a corresponding difference in the DM variation rate. However, for PSR J2055+3829, we find the opposite behavior, where the ingress duration is longer than the egress duration. We suggest that in this case, the cometary-like tail of the outflowing material is swept back, leading to a prolonged ingress phase.
Spider pulsars exhibit a frequency-dependent eclipse duration, often modeled by a single power law~\citep{Polzin2020}. However, wideband observations have shown that for some spider pulsars, such as PSR J1816+4510~\citep{Polzin2020}, a single power law does not accurately describe the eclipse duration at low frequencies, which may arise from a tenuous, swept-back tail of the eclipse medium~\citep{Fruchter1990}.
Our observational bandwidth of 400 MHz is too narrow to thoroughly investigate the frequency dependence of ingress and egress durations. Future observations of PSR J2055+3829 with a wider bandwidth would be valuable for further exploring the properties of the outflowing material from the companion in greater detail.

\section*{Acknowledgements}

This is work is supported by the National Natural Science Foundation of China (No. 12288102, No. 12203092, No. 12041304),  the Major Science and Technology Program of Xinjiang Uygur Autonomous Region (No. 2022A03013-3), the National SKA Program of China (No. 2020SKA0120100), the National Key Research and Development Program of China (No. 2022YFC2205202, No. 2021YFC2203502), the Natural Science Foundation of Xinjiang Uygur Autonomous Region (No. 2022D01B71), the Tianshan Talent Training Program for Young Elite Scientists (No. 2023TSYCQNTJ0024). 
This work made use of the data from the Five-hundred-meter Aperture Spherical radio Telescope, which is a Chinese national megascience facility, operated by National Astronomical Observatories, Chinese Academy of Sciences. The research is partly supported by the Operation, Maintenance and Upgrading Fund for Astronomical Telescopes and Facility Instruments, budgeted from the Ministry of Finance of China (MOF) and administrated by the Chinese Academy of Sciences (CAS).

\section*{Data Availability}
Origin raw data are published by the FAST data center and can be accessed through them. For other intermediate-process data, please contact the author.



\bibliographystyle{mnras}
\bibliography{ms} 



\bsp	
\label{lastpage}
\end{document}